\documentclass[aps,prd,onecolumn,groupedaddress,showpacs,nofootinbib,amssymb]{revtex4}
\usepackage[dvips]{graphicx}
\usepackage{amssymb}
\usepackage{amsmath}
\usepackage{graphicx,,color}
\usepackage{amsfonts}
\usepackage{bm}
\usepackage{cancel}
\usepackage{comment}

\newcommand\be{\begin{equation}}
\newcommand\ee{\end{equation}}

\allowdisplaybreaks[4]

\begin{document}

\title{Inflationary Attractors in $F(R)$ Gravity}
\author{S.D. Odintsov,$^{1,2}$\,\thanks{odintsov@ice.cat}
V.K. Oikonomou,$^{3,4,5}$\,\thanks{v.k.oikonomou1979@gmail.com}}
\affiliation{$^{1)}$ ICREA, Passeig Luis Companys, 23, 08010 Barcelona, Spain\\
$^{2)}$ Institute of Space Sciences (IEEC-CSIC) C. Can Magrans
s/n,
08193 Barcelona, Spain\\
$^{3)}$Department of Physics, Aristotle University of Thessaloniki, Thessaloniki 54124, Greece\\
$^{4)}$ Laboratory for Theoretical Cosmology, Tomsk State
University of Control Systems
and Radioelectronics, 634050 Tomsk, Russia (TUSUR)\\
$^{5)}$ Tomsk State Pedagogical University, 634061 Tomsk, Russia }

\tolerance=5000

\begin{abstract}
In this letter we shall demonstrate that the viable $F(R)$
gravities can be classified mainly into two classes of
inflationary attractors, either the $R^2$ attractors or the
$\alpha$-attractors. To show this, we shall derive the most
general relation between the tensor-to-scalar ratio $r$ and the
spectral index of primordial curvature perturbations $n_s$, namely
the $r-n_s$ relation, by assuming that the slow-roll condition
constrains the values of the slow-roll indices. As we show, the
relation between the tensor-to-scalar ratio and the spectral index
of the primordial curvature perturbations has the form $r\simeq
\frac{48 (1-n_s)^2}{(4-x)^2}$, where the dimensionless parameter
$x$ contains higher derivatives of the $F(R)$ gravity function
with respect to the Ricci scalar, and it is a function of the
$e$-foldings number $N$ and may also be a function of the free
parameters of the various $F(R)$ gravity models. For $F(R)$
gravities which have a spectral index compatible with the
observational data and also yield $x\ll 1$, these belong to the
$R^2$-type of attractors, with $r\sim 3 (1-n_s)^2$, and these are
viable theories. Moreover, in the case that $x$ takes larger
values in specific ranges and is constant for a given $F(R)$
gravity, the resulting $r-n_s$ relation has the form $r\sim 3
\alpha (1-n_s)^2$, where $\alpha$ is a constant. Thus we conclude
that the viable $F(R)$ gravities may be classified into two
limiting types of $r-n_s$ relations, one identical to the $R^2$
model at leading order in $x$, and one similar to the
$\alpha$-attractors $r-n_s$ relation, for the $F(R)$ gravity
models that yield $x$ constant. Finally, we also discuss the case
that $x$ is not constant.
\end{abstract}


\maketitle

\section{Introduction}

The inflationary era is one of the theoretical proposals that
describe the post-Planckian epoch of our Universe. At the
beginning of this primordial epoch, the Universe can be described
in a classical way, and is four dimensional, at least in the most
inflationary proposals. The inflationary scenario
\cite{inflation1,inflation2,inflation3,inflation4} in its various
versions solves most of the prominent problems of the Big Bang
cosmology, such as the flatness and horizon problems, but still
even up to date, it is not certain whether this primordial epoch
even occurred. Only the direct detection of $B$-modes in the
Cosmic Microwave Background may verify the inflationary era
\cite{Kamionkowski:2015yta}.

The standard approach in theoretical cosmology that describes the
inflationary epoch is to use scalar fields, however another
successful description is to use the modified gravity description
of inflation. Many theoretical models of modified gravity may
successfully describe the inflationary era, see for example the
reviews \cite{reviews1,reviews2,reviews3,reviews4}, with $F(R)$
gravity theories playing a prominent role among all inflationary
models of modified gravity. The latest Planck observational data
further restricted the characteristics of the primordial era, and
have narrowed down the number of viable modified gravity theories.
Since we are living in the precision cosmology era, a theoretical
model for the primordial era, must pass a number of tests in order
for it to be considered as successful.

In modified gravity, the most important and crucial tests that a
model must pass, is the production of a nearly scale invariant
power spectrum of primordial curvature perturbations, and also to
produce a significantly small tensor-to-scalar perturbations
ratio. The procedure for calculating these two observational
indices of inflation is quite tedious, due to the fact that the
modified gravity gravitational equations are in most cases hard to
solve analytically. A crucial assumption for the inflationary era
is the slow-roll assumption, which simplifies to a great extent
the calculations of the inflationary indices. Then, the latter are
expressed in terms of the $e$-foldings number, and the various
model parameters, and a direct confrontation of the theoretical
proposal with the observational data may be done. However, even in
the slow-roll approximation, many simplifications are required in
order for a result to be obtained in closed form. Thus, several
techniques that may reveal universality properties of various
models, or phase space techniques
\cite{Odintsov:2017tbc,Bahamonde:2017ize} may offer valuable
insights on the inflationary predictions of a model or even a
class of models. For example, the $\alpha$-attractor models
\cite{alpha1,alpha2,alpha3,alpha4,alpha5,alpha6,alpha7,alpha8,alpha9,alpha10,alpha10a,alpha11,alpha12,vernov}
are a characteristic example of models that have a universal
behavior, quantified in the spectral index and tensor-to-scalar
ratio behavior as functions of the $e$-foldings number. Also
several string theory models also have this sort of behavior
\cite{Burgess:2016owb}, so this is a clear indication of an
underlying theme yet to be understood, which relates phenomenally
distinct theoretical models.

In this line of research, in this letter, we shall study the
tensor-to-scalar $r$ and spectral index $n_s$ relation, to which
we shall refer as $r-n_s$ relation, for $F(R)$ gravity theories in
vacuum. Particularly, by assuming that the slow-roll conditions
apply for the Hubble rate and its derivatives, we shall derive a
general expression for the $r-n_s$ relation, that as we will show
will have the form $r\simeq \frac{48 (1-n_s)^2}{(4-x)^2}$, with
the dimensionless parameter $x$ being related to higher
derivatives of the $F(R)$ gravity function with respect to the
Ricci scalar. The parameter $x$ is in general a function of the
$e$-foldings number $N$, and may also be a function of the free
parameters of the various $F(R)$ models. Our focus will be on
$F(R)$ gravities that result to a spectral index compatible with
the observational data, and also yield $x\ll 1$. As we show these
belong to a class of models that the $r-n_s$ relation has the form
$r\sim 3 (1-n_s)^2$, at leading order in $x$, which we shall call
$R^2$-attractors, since the $R^2$ inflation
\cite{Starobinsky:1980te} (see also \cite{Bezrukov:2007ep} for the
Jordan frame non-minimally coupled scalar analogue) also has
exactly this $r-n_s$ relation. Furthermore, if $x$ takes larger
values in specific ranges, and also is constant, the resulting
$r-n_s$ relation has the form $r\sim 3 \alpha (1-n_s)^2$, where
$\alpha$ is a constant. This $r-n_s$ relation is of the
$\alpha$-attractor form, so our results indicate that the viable
$F(R)$ gravities may be classified mainly into two limiting types
of $r-n_s$ relations, one of which is identical to the $R^2$ model
at leading order in $x$, and the other is identical to the
$\alpha$-attractors $r-n_s$ relation. Finally, we also discuss the
case for which $x$ takes constant values and does not satisfy the
constraint $x\ll 1$ anymore.

In the rest of this letter, the geometric background will be
assumed to be a flat Friedmann-Robertson-Walker (FRW) metric of
the form,
\begin{equation}
\label{JGRG14} ds^2 = - dt^2 + a(t)^2 \sum_{i=1,2,3}
\left(dx^i\right)^2\, ,
\end{equation}
with $a(t)$ being the scale factor.

\section{$F(R)$ Gravity Inflation, $R^2$ Attractors and $\alpha$-Attractors}

We shall consider $F(R)$ gravity theory in vacuum, so the
gravitational action is,
\begin{equation}\label{action1dse}
\mathcal{S}=\frac{1}{2\kappa^2}\int \mathrm{d}^4x\sqrt{-g}F(R),
\end{equation}
with $\kappa^2$ standing for $\kappa^2=8\pi G=\frac{1}{M_p^2}$ and
with $M_p$ being the reduced Planck mass. In the metric formalism,
the gravitational equations of motion are obtained by varying the
action with respect to the metric tensor, so these are,
\begin{equation}\label{eqnmotion}
F_R(R)R_{\mu \nu}(g)-\frac{1}{2}F(R)g_{\mu
\nu}-\nabla_{\mu}\nabla_{\nu}F_R(R)+g_{\mu \nu}\square F_R(R)=0\,
,
\end{equation}
where $F_R=\frac{\mathrm{d}F}{\mathrm{d}R}$. Upon rewriting Eq.
(\ref{eqnmotion}) we obtain,
\begin{align}\label{modifiedeinsteineqns}
R_{\mu \nu}-\frac{1}{2}Rg_{\mu
\nu}=\frac{\kappa^2}{F_R(R)}\Big{(}T_{\mu
\nu}+\frac{1}{\kappa^2}\Big{(}\frac{F(R)-RF_R(R)}{2}g_{\mu
\nu}+\nabla_{\mu}\nabla_{\nu}F_R(R)-g_{\mu \nu}\square
F_R(R)\Big{)}\Big{)}\, .
\end{align}
The gravitational equations of motion for the FRW metric
(\ref{JGRG14}) become,
\begin{align}
\label{JGRG15} 0 =& -\frac{F(R)}{2} + 3\left(H^2 + \dot H\right)
F_R(R) - 18 \left( 4H^2 \dot H + H \ddot H\right) F_{RR}(R)\, ,\\
\label{Cr4b} 0 =& \frac{F(R)}{2} - \left(\dot H +
3H^2\right)F_R(R) + 6 \left( 8H^2 \dot H + 4 {\dot H}^2 + 6 H
\ddot H + \dddot H\right) F_{RR}(R) + 36\left( 4H\dot H + \ddot
H\right)^2 F_{RRR} \, ,
\end{align}
with $F_{RR}=\frac{\mathrm{d}^2F}{\mathrm{d}R^2}$, and
$F_{RRR}=\frac{\mathrm{d}^3F}{\mathrm{d}R^3}$, and in addition,
$H$ is the Hubble rate $H=\dot a/a$ while the Ricci scalar for the
FRW metric is $R=12H^2 + 6\dot H$.

We shall be interested in deriving a general expression for the
functional relation of the tensor-to-scalar and spectral index of
the primordial scalar perturbations for a general $F(R)$ gravity,
to which we shall refer for simplicity as $r-n_s$ relation. To
this end, we shall use the slow-roll indices, and we shall assume
that the slow-roll approximation holds true, that is,
\begin{equation}\label{slowrollconditionshubble}
\ddot{H}\ll H\dot{H},\,\,\, \frac{\dot{H}}{H^2}\ll 1\, .
\end{equation}
The slow-roll indices, namely $\epsilon_1$ ,$\epsilon_2$,
$\epsilon_3$, $\epsilon_4$, quantify the dynamics of the
inflationary era, and the general expression for these is
\cite{Hwang:2005hb,reviews1},
\begin{equation}
\label{restofparametersfr}\epsilon_1=-\frac{\dot{H}}{H^2}, \quad
\epsilon_2=0\, ,\quad \epsilon_3= \frac{\dot{F}_R}{2HF_R}\, ,\quad
\epsilon_4=\frac{\ddot{F}_R}{H\dot{F}_R}\,
 .
\end{equation}
A crucial assumption for our study is that the slow-roll indices
satisfy the slow-roll condition $\epsilon_i\ll 1$, $i=1,3,4$.
Without the slow-roll assumption, the spectral index and the
tensor-to-scalar ratio in terms of the slow-roll indices are
\cite{reviews1,Hwang:2005hb},
\begin{equation}
\label{epsilonall} n_s=
1-\frac{4\epsilon_1-2\epsilon_3+2\epsilon_4}{1-\epsilon_1},\quad
r=48\frac{\epsilon_3^2}{(1+\epsilon_3)^2}\, .
\end{equation}
The authors of Ref. \cite{Hwang:2005hb} derived the expression for
the spectral index appearing in Eq. (\ref{epsilonall}) by using
the condition $\dot{\epsilon}_1=0$, however, in Ref.
\cite{Oikonomou:2020krq} we showed that the condition
$\dot{\epsilon}_1=0$ is in fact superfluous and even misleading.
In fact, the same expression for the spectral index as in Eq.
(\ref{epsilonall}) can be derived without assuming
$\dot{\epsilon}_1=0$, as we showed in Ref.
\cite{Oikonomou:2020krq}.

The expression for the tensor-to-scalar ratio is derived from the
power spectrum as the fraction of the tensor perturbation $P_T$
and the scalar perturbation $P_S$ as follows,
\begin{equation}\label{tensorananalytic}
r=\frac{P_T}{P_S}=8 \kappa^2 \frac{Q_s}{F_R}\, ,
\end{equation}
where,
\begin{equation}
\label{qsfrpreliminary}
Q_s=\frac{3\dot{F_R}^2}{2F_RH^2\kappa^2(1+\epsilon_3)^2}\, .
\end{equation}
So by combining Eqs. (\ref{tensorananalytic}) and
(\ref{qsfrpreliminary}) we have,
\begin{equation}\label{ranalyticfinal}
r=48 \frac{\dot{F_R}^2}{4F_R^2H^2(1+\epsilon_3)^2}\, ,
\end{equation}
and recalling that $\epsilon_3= \frac{\dot{F}_R}{2HF_R}$, we
finally have,
\begin{equation}\label{ranalyticfinal1}
r=48\frac{\epsilon_3^2}{(1+\epsilon_3)^2}\, .
\end{equation}
Let us now take into account that the slow-roll indices satisfy
the slow-roll condition $\epsilon_i\ll 1$, $i=1,3,4$. From the
Raychaudhuri equation, without however assuming the slow-roll
conditions, we have,
\begin{equation}\label{approx1}
\epsilon_1=-\epsilon_3(1-\epsilon_4)\, ,
\end{equation}
so under the slow-roll condition we have $\epsilon_1\simeq
-\epsilon_3$, and therefore, the spectral index becomes,
\begin{equation}
\label{spectralfinal} n_s\simeq 1-6\epsilon_1-2\epsilon_4\, ,
\end{equation}
while the tensor-to-scalar ratio becomes $r\simeq 48
\epsilon_3^2$, and since $\epsilon_1\simeq -\epsilon_3$
\begin{equation}
\label{tensorfinal} r\simeq 48\epsilon_1^2\, .
\end{equation}
Let us now focus on the slow-roll index
$\epsilon_4=\frac{\ddot{F}_R}{H\dot{F}_R}$ and we shall try to
express it as a function of $\epsilon_1$. We easily find that,
\begin{equation}\label{epsilon41}
\epsilon_4=\frac{\ddot{F}_R}{H\dot{F}_R}=\frac{\frac{d}{d
t}\left(F_{RR}\dot{R}\right)}{HF_{RR}\dot{R}}=\frac{F_{RRR}\dot{R}^2+F_{RR}\frac{d
(\dot{R})}{d t}}{HF_{RR}\dot{R}}\, .
\end{equation}
However, $\dot{R}$ is equal to,
\begin{equation}\label{rdot}
\dot{R}=24\dot{H}H+6\ddot{H}\simeq 24H\dot{H}=-24H^3\epsilon_1\, ,
\end{equation}
where we used the slow-roll approximation condition $\ddot{H}\ll H
\dot{H}$. Note that if we apply the derivative with respect to the
cosmic time in Eq. (\ref{epsilon41}), a term $\sim
\frac{d^3H}{dt^2}$ would appear, however, before we applying the
derivative in Eq. (\ref{epsilon41}), we assumed in Eq.
(\ref{rdot}) that $\dot{R}\simeq 24H\dot{H}\simeq
-24H^3\epsilon_1$, so we avoided the appearance of the term $\sim
\frac{d^3H}{dt^2}$. Hence in our case, the only condition required
in our case is $\ddot{H}\ll H \dot{H}$.

By substituting Eq. (\ref{rdot}) in Eq. (\ref{epsilon41}) we get
after some algebra,
\begin{equation}\label{epsilon4final}
\epsilon_4\simeq -\frac{24
F_{RRR}H^2}{F_{RR}}\epsilon_1-3\epsilon_1+\frac{\dot{\epsilon}_1}{H\epsilon_1}\,
.
\end{equation}
But the term $\dot{\epsilon}_1$ reads,
\begin{equation}\label{epsilon1newfiles}
\dot{\epsilon}_1=-\frac{\ddot{H}H^2-2\dot{H}^2H}{H^4}=-\frac{\ddot{H}}{H^2}+\frac{2\dot{H}^2}{H^3}\simeq
2H \epsilon_1^2\, ,
\end{equation}
therefore the final approximate expression of $\epsilon_4$ is,
\begin{equation}\label{finalapproxepsilon4}
\epsilon_4\simeq -\frac{24
F_{RRR}H^2}{F_{RR}}\epsilon_1-\epsilon_1\, .
\end{equation}
We introduce the dimensionless parameter $x$ which is defined as
follows,
\begin{equation}\label{parameterx}
x=\frac{48 F_{RRR}H^2}{F_{RR}}\, ,
\end{equation}
and in terms of it, the parameter $\epsilon_4$ reads,
\begin{equation}\label{epsilon4finalnew}
\epsilon_4\simeq -\frac{x}{2}\epsilon_1-\epsilon_1\, .
\end{equation}
By substituting $\epsilon_4$ from Eq. (\ref{epsilon4finalnew}) in
Eq. (\ref{spectralfinal}), the spectral index can be written as a
function of $\epsilon_1$,
\begin{equation}\label{asxeto1}
n_s-1=-4\epsilon_1+x\epsilon_1\, ,
\end{equation}
so by solving the above with respect to $\epsilon_1$ we get,
\begin{equation}\label{spectralasfunctionofepsilon1}
\epsilon_1=\frac{1-n_s}{4-x}\, ,
\end{equation}
and by substituting in the final expression for the
tensor-to-scalar ratio in Eq. (\ref{tensorfinal}), we get,
\begin{equation}\label{mainequation}
r\simeq \frac{48 (1-n_s)^2}{(4-x)^2}\, .
\end{equation}
The above equation is the main result  of this letter, and in the
rest of this work we shall discuss the various implications of Eq.
(\ref{mainequation}) depending on the values of the parameter $x$.

Firstly let us note that the parameter $x$ is not constant for
general $F(R)$ gravities, although in some cases can be constant.
In general it depends on the $e$-foldings number, and can be
calculated if the Friedmann equation is solved for the
corresponding vacuum $F(R)$ gravity. This is not an easy task for
most of the $F(R)$ gravities, even in the slow-roll approximation,
so let us make some estimations about the values of $x$. Its
explicit form is hard to find analytically, but we can make some
general estimations about the values of it, and of course discuss
which values it can take in several limiting cases. Recall from
Eq. (\ref{parameterx}) that $x=\frac{48 F_{RRR}H^2}{F_{RR}}$, so
it is obvious that $x$ is a function of the Hubble rate and its
higher derivatives. If the Friedmann equation is solved for a
given $F(R)$ gravity, and the Hubble rate is found, then by using
the definition of the $e$-foldings number $N=\int_{t_i}^{t_f}H(t)d
t$ and by inverting it, we can express the horizon crossing time
instance $t_i$ as a function of the $e$-foldings number and the
time instance that inflation ends $t_f$, that is $t_i=t_i(N,t_f)$.
The time instance $t_f$ can be found by equating
$\epsilon_1(t_f)=1$, which is the condition when inflation ends.
In effect, by using the Hubble rate solution for a specific model
and the relation $t_i=t_i(N,t_f)$, we can express the parameter
$x$ as a function of the $e$-foldings number and of the set of the
free parameters of the model, which we denote as $\xi$, so we have
$x=x(N,\xi)$.

Now there are three possibilities, firstly the case that $x\ll 1$
(or $x\simeq 0$ equivalently), secondly that $x\sim
\mathcal{O}(4)$ and thirdly that $x\gg 1$. These three
possibilities are determined by the behavior of the function
$x=x(N,\xi)$ for large $N$ (recall $N\sim 60$ for a sufficiently
long inflationary era), and of course the behavior is also
determined by the free parameters. But in any case, these three
possibilities always occur, and this behavior covers also the
cases that $x$ is a constant number.

Let us discuss the three cases separately. If $x\ll 1$ the $r-n_s$
relation becomes approximately,
\begin{equation}\label{rnsstarobinsky}
r\sim 3 (1-n_s)^2\, ,
\end{equation}
which is the $r-n_s$ relation obeyed by the $R^2$ model (in which
case $x=0$). The same $r-n_s$ relation is obtained if $x=0$
directly. In fact when $x\ll 1$, the relation of Eq.
(\ref{rnsstarobinsky}) is the leading order result, since by
expanding relation (\ref{mainequation}) for $x\ll 1$, we get,
\begin{equation}\label{taylor}
r\simeq  3 (1 - n_s)^2+\frac{3 (1 - n_s)^2}{2}x+\frac{9 (1 -
n_s)^2}{16}x^2\, .
\end{equation}
So it is apparent that the $r-n_s$ model relation of the $R^2$
model is the leading order term in the expansion (\ref{taylor}).
In effect, all theories that yield $x\sim 0$ or $x\ll 1$, result
to the same $r-n_s$ relation that the $R^2$ model has, at leading
order in $x$. Note that in this case we did not specify whether
the parameter $x$ is positive or negative, because the sign of $x$
does not affect at all the final $r-n_s$ relation at leading
order, when $x\ll 1$. The case $x\sim \mathcal{O}(4)$ is more
involved, since the tensor to scalar ratio may actually blow up
for $x\to 4$. This is more model dependent and $F(R)$ gravities
which yield $x\sim 4$ certainly are non-viable. However, certain
values of $x$ with $x$ being in some interval around $4$ may yield
viability. For these theories, the $r-n_s$ would be of the form
$r\simeq \frac{48 (1-n_s)^2}{(4-x)^2}$. Now in the case that $x\gg
1$, the $r-n_s$ relation (\ref{mainequation}) can be approximated,
\begin{equation}\label{mainequationrevisionapprox1}
r\simeq \frac{48 (1-n_s)^2}{x^2}\, .
\end{equation}
One may argue that since the above relation is $N$-dependent and
also depends on the free parameters of the model, and since $n_s$
is within the observational limits, then due to the fact that
$x\gg 1$, this means that $x$ must contain positive powers of $N$,
hence $r$ is predicted to be very small in magnitude. However,
this is not true, since recall that from Eq.
(\ref{epsilon4finalnew}) $\epsilon_4\sim
-\frac{x}{2}\epsilon_1-\epsilon_1$, hence if $x\gg 1$, then the
slow-roll condition on the slow-roll index $\epsilon_4$ would no
longer be valid. Hence, the $x\gg 1$ case seems to produce a break
down of the slow-roll condition, and seems model dependent. Note
that if $x<0$, the same results apply in the case that $|x|\gg 1$,
since in this case,
\begin{equation}\label{mainequationrevisionapprox12}
r\simeq \frac{48 (1-n_s)^2}{(-|x|)^2}=\frac{48 (1-n_s)^2}{x^2}\, .
\end{equation}

The same arguments apply for the case that $x$ is a constant, but
of course in this case, there is no $N$-dependence in the $r-n_s$
relation. In fact, the $r-n_s$ relation takes the following form
for constant $x$,
\begin{equation}\label{alphaattractor}
r\sim 3\alpha (1-n_s)^2\, ,
\end{equation}
where $\alpha=\frac{16}{(4-x)^2}$, which is identical, at least
functionally, to the $\alpha$-attractors relation \cite{alpha6},
and is also related to several types of string theory $r-n_s$
relations \cite{Burgess:2016owb}. The $\alpha$-attractor models
belong to large class of minimally coupled scalar theories and
non-minimally coupled theories
\cite{alpha1,alpha2,alpha3,alpha4,alpha5,alpha6,alpha7,alpha8,alpha9,alpha10,alpha10a,alpha11,alpha12,vernov}.
Particularly, canonical scalar field models like the Starobinsky
model and the Higgs model are some limiting cases of the
$\alpha$-attractors class of models. The fact that the
$\alpha$-attractor models result to the same $r-n_s$ relation
indicates that there has to be a common origin for all these
models. Particularly, when seen from the scalar field point of
view, in the presence of a potential, the potential itself for
$\alpha$-attractors has a large plateau for large scalar field
values, and most of these models are asymptotically similar to the
hybrid inflation scenario \cite{hybrid}. It is noteworthy to say
that some of the $\alpha$-attractor models have a supergravity
origin \cite{susybr1}, with the most interesting feature being the
fact that the minimum of the scalar potential is exactly the point
where supersymmetry breaking occurs.

The functional similarity of the $x=$constant case with the
$\alpha$-attractors, quantified by relation (\ref{alphaattractor})
does not necessarily imply that the model is viable. If $x\ll 1$
or $x\sim 0$, then Eq. (\ref{alphaattractor}) reduces to Eq.
(\ref{rnsstarobinsky}), and if $x\gg 1$, Eq.
(\ref{alphaattractor}) reduces to Eq.
(\ref{mainequationrevisionapprox1}) and as we saw only the $x\ll
1$ case leads to viable results. However, if $x\sim 4$, then the
tensor-to-scalar ratio might take significantly large values, so a
non-viable theory might be obtained. Nevertheless if $x$ takes
constant values around the value 4, then there is the possibility
that the theory might be viable, with the $r-n_s$ relation in this
case being of the $\alpha$-attractors form given in Eq.
(\ref{alphaattractor}), namely $r\sim 3\alpha (1-n_s)^2$, with
$\alpha=\frac{16}{(4-x)^2}$. But of course all these results are
strongly model dependent and hold true only for those $F(R)$
gravities which satisfy $x=$constant constraint.

In order to have a more clear idea of the $x=$constant case, we
shall perform some qualitative analysis by using some numerical
values. To start off, recall that in Eq. (\ref{epsilon4finalnew}),
the slow-roll parameter is $\epsilon_4\sim
-\frac{x}{2}\epsilon_1-\epsilon_1$. By assuming that the slow-roll
condition applies, then $\epsilon_4$ can be at most
$\epsilon_4\sim \mathcal{O}(10^{-2})$, and according to the 2018
Planck constraints on the slow-roll index $\epsilon_1$
\cite{Akrami:2018odb}, the latter is of the order $\epsilon_1\sim
\mathcal{O}(10^{-3})$, so $x$ can at most be of the order $x\sim
\mathcal{O}(10^{1})$, assuming that $x$=constant. The latest
Planck data indicate that the spectral index and the
tensor-to-scalar ratio are constrained as follows
\cite{Akrami:2018odb},
\begin{equation}\label{observationaldatanewresults}
n_s=0.962514\pm 0.00406408,\,\,\,r<0.064\, .
\end{equation}
So if an $F(R)$ theory yields a viable spectral index $n_s$, the
corresponding parameter $x$ must take the following limiting
values in order to render the tensor-to-scalar ratio $r$ also
viable: if $n_s=0.9691$ (maximum allowed value by Planck 2018
data), then $x$ must not be in the range $3.15377<x<4.84623$ and
if $n_s=0.9607$ (minimum allowed value by Planck 2018 data), then
$x$ must not be in the range $2.92373<x<5.07627$. If $x$ belongs
in the above ranges, for $N\sim 50-60$, the theory is not viable.

\begin{figure}[h!]
\centering
\includegraphics[width=18pc]{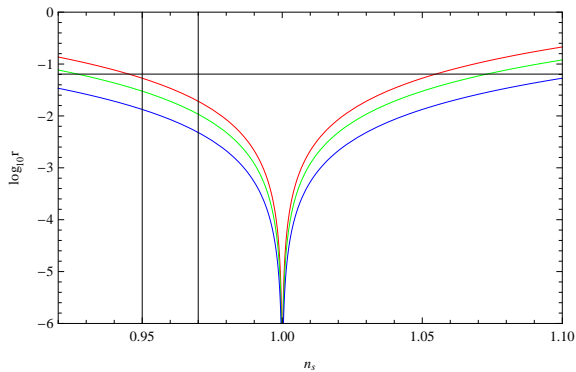}
\caption{Plots of the $\log_{10}r-n_s$ relation for $x=5.5$ (red),
$x=6$ (green), $x=7$ (blue), for $n_s=0.9691$. The horizontal
black line indicates the Planck 2018 upper limit for the
$\log_{10} r$, and the vertical black lines, the allowed range of
the spectral index values from Planck 2018. In between these
ranges, the $F(R)$ models are viable.} \label{plot1}
\end{figure}

Coming back to the problem at hand, with regard to the values of
$x$ and the $\alpha$-attractor behavior, if for example if
$n_s=0.9691$, which recall that is the maximum allowed value by
Planck 2018 data, then $x$ must be chosen to be $x>4.84623$ or
$x<3.15377$, and at the same time it should not be too large in
order for the slow-roll condition to hold true. In Fig.
\ref{plot1} we present the plots of the $\log_{10}r-n_s$ relation
for $x=5.5$ (red), $x=6$ (green), $x=7$ (blue), for $n_s=0.9691$.
The horizontal black line indicates the Planck 2018 upper limit
for the $\log_{10} r$, while the vertical black lines, the allowed
range of the spectral index values from Planck 2018.
Let us note that for this simple qualitative analysis we assumed
that $x$ is constant.

\section{Conclusions}

Thus what we demonstrated is that the viable $F(R)$ gravities may
be classified mainly into two limiting types of $r-n_s$ relations,
one identical to the $R^2$ model at leading order in $x$, when
$x\ll 1$, and one identical to the $\alpha$-attractors $r-n_s$
relation, when $x$ takes constant values for a given $F(R)$
gravity. However, we need to note that we did not consider cases
for which one of the slow-roll indices may not comply with the
slow-roll assumptions, like for instance in the constant-roll
case. This would require a different treatment of the study we
performed in this letter, since $\ddot{H}\sim H\dot{H}$ in the
constant-roll case.

It is noteworthy to state that such a classification among some
viable $F(R)$ gravity theories might also be justified by the
conformal invariance of the observational indices, so our result
is the Jordan frame manifestation of the Einstein frame
$\alpha$-attractors result, since the spectral index $n_s$ and the
tensor-to-scalar ratio $r$ turn out to be nearly the same if
calculated in the Jordan or in the Einstein frame
\cite{Kaiser:1995nv,Domenech:2016yxd,Brooker:2016oqa,Kaiser:1994vs}.

Furthermore, most of $F(R)$ gravity approaches use
$\dot{\epsilon}_i=0$, $i=1,2,3$, for extracting the spectral index
of Eq. (\ref{epsilonall}). This however is a severe inconsistency
in the formalism of $F(R)$ gravity inflation. Actually this
condition used in Ref. \cite{Hwang:2005hb}, namely
$\dot{\epsilon}_1=0$, is too strong, since the only thing required
for obtaining the spectral index of Eq. (\ref{epsilonall}), is
$\dot{\epsilon}_i\ll 1$, $i=1,2,3$. We resolved this issue in
detail in Ref. \cite{Oikonomou:2020krq}. This slight change of the
condition on the derivatives of the slow-roll indices, and
specifically of the slow-roll index $\epsilon_1$ from
$\dot{\epsilon}_1=0$ to $\dot{\epsilon}_1\ll 1$, has an impact on
$\epsilon_4$ and subsequently on the $r-n_s$ relation. Moreover it
can have some effect on the phenomenology of some $F(R)$
gravities. This issue is quite intriguing, since the condition
$\dot{\epsilon}_1=0$ would make even viable $F(R)$ gravities to
yield non-viable results, as shown in \cite{Oikonomou:2020krq}
where we showed explicitly how the condition $\dot{\epsilon}_1=0$
would render even viable $F(R)$ gravities, like the $R^2$ model
non-viable. In fact, the authors of Ref. \cite{Hwang:2005hb}
derived the expression for the spectral index we quoted in Eq.
(\ref{epsilonall}), by using the condition $\dot{\epsilon}_1=0$,
which is a crucial inconsistency, which we resolved in a
mathematical way in Ref. \cite{Oikonomou:2020krq}. As we showed in
Ref. \cite{Oikonomou:2020krq}, the result of Ref.
\cite{Hwang:2005hb} which we used in Eq. (\ref{epsilonall}) in the
present letter, holds true, without the condition
$\dot{\epsilon}_1=0$. If the condition $\dot{\epsilon}_1=0$ is
required, then the expression for the spectral index given in Eq.
(\ref{epsilonall}) of the present draft would only hold true for
power-law gravities $F(R)\sim R^n$, for $n\neq 2$, which obviously
does not include the $R^2$ model. In addition, if the condition
$\dot{\epsilon}_1=0$ is assumed to hold true, then the $R^2$ model
would yield an exactly scale invariant power spectrum, see Ref.
\cite{Oikonomou:2020krq} for details. Finally, let us note that
the formalism and results which we developed in this work, does
not apply in the case that the $F(R)$ gravity model yields
$\dot{\epsilon}_1= 0$, and in effect some power-law $F(R)$
gravities of the form $F(R)=R+\gamma R^n$, with $n\neq 2$ cannot
be appropriately described by our formalism. We are working along
this research lines in order to explain the problem with these
theories, strongly related to the slow-roll assumption, and we
address this issue in a future work.

\section*{Acknowledgments}

This work is supported by MINECO (Spain), FIS2016-76363-P, and by
project 2017 SGR247 (AGAUR, Catalonia) (S.D.O).

\end{document}